\documentclass[fleqn,usenatbib]{mnras}

\usepackage{newtxtext,newtxmath}

\usepackage[T1]{fontenc}

\DeclareRobustCommand{\VAN}[3]{#2}
\let\VANthebibliography\thebibliography
\def\thebibliography{\DeclareRobustCommand{\VAN}[3]{##3}\VANthebibliography}

\usepackage{graphicx}	
\usepackage{amsmath}	

\title[Electrostatic Dissolution of Planetesimals]{The Dissolution of Planetesimals in Electrostatic Fields}

\author[F. C. Onyeagusi et al.]{
F. C. Onyeagusi,$^{1}$\thanks{E-mail: florence.onyeagusi@uni-due.de (FCO)}
J. Teiser,$^{1}$
T. Becker$^{1}$
and G. Wurm$^{1}$
\\
$^{1}$University of Duisburg-Essen, Faculty of Physics,
Lotharstr. 1, 47057 Duisburg, Germany}

\date{Accepted XXX. Received YYY; in original form ZZZ}

\pubyear{2023}

\begin{document}
\label{firstpage}
\pagerange{\pageref{firstpage}--\pageref{lastpage}}
\maketitle

\begin{abstract}
Planetesimals or smaller bodies in protoplanetary disks are often considered to form as pebble piles in current planet formation models. They are supposed to be large but loose, weakly bound clusters of more robust dust aggregates. This makes them easy prey for destructive processes. In microgravity experiments, we apply strong electric fields on clusters of slightly conductive dust aggregates. We find that this generates enough tensile stress on the fragile clusters to sequentially rip off the aggregates from the cluster. 
These experiments imply that electric fields in protoplanetary disks can dissolve pebble pile planetesimals. This process might induce a bias for the local planetesimal reservoir in regions with strong fields. Planetesimals prevail with certain kinds of compositions where they are either good isolators or compacted bodies. The less lucky ones generate pebble clouds which might be observable as signposts of electrostatic activity in protoplanetary disks.
\end{abstract}




\section{Introduction}

Dust in protoplanetary disks grows through various phases. Starting as micron-sized dust, it easily forms millimeter-sized aggregates which become compact, eventually \citep{Blum2008, Weidling2009, Kelling2014, Kruss2017, WurmTeiser2021}. These aggregates can then form larger clusters, e.g. moderated by triboelectric charging, be concentrated and gently collapse into planetesimals \citep{Lee2015, Steinpilz2020a, Schneider2019, Teiser2021, Youdin2005, Johansen2007, Johansen2014}. In the end, this is supposed to produce planetesimals which are clusters of very weakly bound dust aggregates, even if these clusters are km-sized \citep{Blum2017, Visser2021}.

With the initial bouncing barrier for mm-sized aggregates \citep{Zsom2010, Kelling2014, Demirci2017}, it is very promising that there are mechanisms that are proposed to support further growth and it is tempting to assume that there is only a way forward to larger bodies with the need to form planets in mind. However, the contacts between the pebbles are not strong which is a general weakness for these piles of aggregates and there are efficient ways to undo any foregone growth. 

Catastrophic collisions are a prominent representative of an efficient destruction process \citep{Dobinson2016}. It does not have to be that violent though. Just the headwinds these bodies encounter while moving through the gas of a protoplanetary disk can easily erode the whole body away by regular gas drag in the inner parts of the disk \citep{Paraskov2006, Demirci2019, Demirci2020b, Demirci2020a, Schoenau2023, Schaffer2020, Cedenblad2021, Rozner2020, Grishin2020, Quillen2023}. We put emphasis on this gentle process, as erosion turns the object back into a cloud of rather monodisperse pebbles, which might be important for observation.

Along these lines, we introduce another gentle destructive process for pebble pile planetesimals here: electrostatic erosion. Electric fields in protoplanetary disks apply a tensile stress to the body which can rip off the surface particles layer by layer. We will detail this and our supporting experiments below but in short, a planetesimal passing through an electric field might be dissolved into its constituent pebbles. This would provide some constraints on stable planetesimal populations. Maybe even more importantly,  this destruction process might be observable and actually be one of the few ways to trace electric activity in the disk.

Observations of protoplanetary disks have become ever more detailed in recent years and ALMA observations, for example, now provide a wealth of images with substructures of all kinds from rings to spirals to annular structures  \citep{Andrews2018, Huang2018b, Huang2018, Michel2023}. In continuum, these observations trace the emission by dust and sand sized particles. Therefore, these substructures have been modeled also by a wealth of processes that either increase the surface density of solids (locally) or decrease it. The most popular explanation for rings is embedded planets that clear out part of the disk or produce dust by collisional grinding of planetesimals in their wake \citep{Boley2017, Dobinson2016, Ziampras2023}. Spirals have also been proposed to be related to shadows in transition disks \citep{Cuello2019}. In such optically thin settings, photophoretic formation of rings is possible as well \citep{Krauss2005, Husmann2016}. Rings and annular substructures are often more generally attributed to particle traps, i.e. due to vortices or pressure maxima in general which tend to concentrate certain grain sizes \citep{Raettig2021, Vericel2021, Lee2022, Pinilla2015, vanderMarel2015}.
Adding more ideas, rings, gaps and particle concentrations can also be connected to snowlines of various materials \citep{Charnoz2021, Izidoro2022, Saito2011, Ros2013, Johansen2022, Fritscher2021},  the Curie-line \citep{Bogdan2023} or thermal wave instabilities \citep{Ueda2021}.

This list is not exhaustive but the point is, that it becomes visible whenever pebbles have just been formed and have a high surface density or if they are concentrated at certain locations. Observations might be tracing the dominant process that leads to this high concentration, may it be trapping or destruction.
In this sense, we add another potentially observable process here. The interesting prospect is that this might trace strong electric fields. So far, lightning is for example not observable in protoplanetary disks, but our new mechanism detailed below destroys planetesimals in such locations of high electric fields. Therefore, substructures observed in protoplanetary disks might be signposts of lightning or electric field activity.

\section{Microgravity experiments}

The experiments which were the motivation to come up with the electrostatic erosion mechanism were carried out at the drop tower in Bremen. They follow in line with other experiments to study the evolution of clusters of tribocharged grains \citep{Lee2015, Steinpilz2020a, Teiser2021, Jungmann2018, Jungmann2021c, Jungmann2022, Steinpilz2020b}.

Here, we used a Martian dust simulant (MGS) \citep{Cannon2019} as basic sample material.
Aggregates were prepared in the laboratory by vibrating the µm dust sample on a commercial shaker. Individual dust grains stick together due to cohesive forces. This produces (sub-)mm-sized compact dust aggregates which first grow, subsequently compact and finally hit a bouncing limit, quite in analogy to the bouncing barrier in protoplanetary disks \citep{Weidling2009, Onyeagusi2023a, Onyeagusi2023b, Zsom2010, Kruss2016}. The aggregates that are created this way are irregularly shaped and are sifted into the size ranges of 150 - 500~µm and 500~µm - 2.0~mm.
These particles were placed within an experimental chamber of 10\,cm x 5\,cm x 5\,cm. 

\subsection{Drop tower setup}

\begin{figure}
   \centering
      \includegraphics[width = \linewidth]{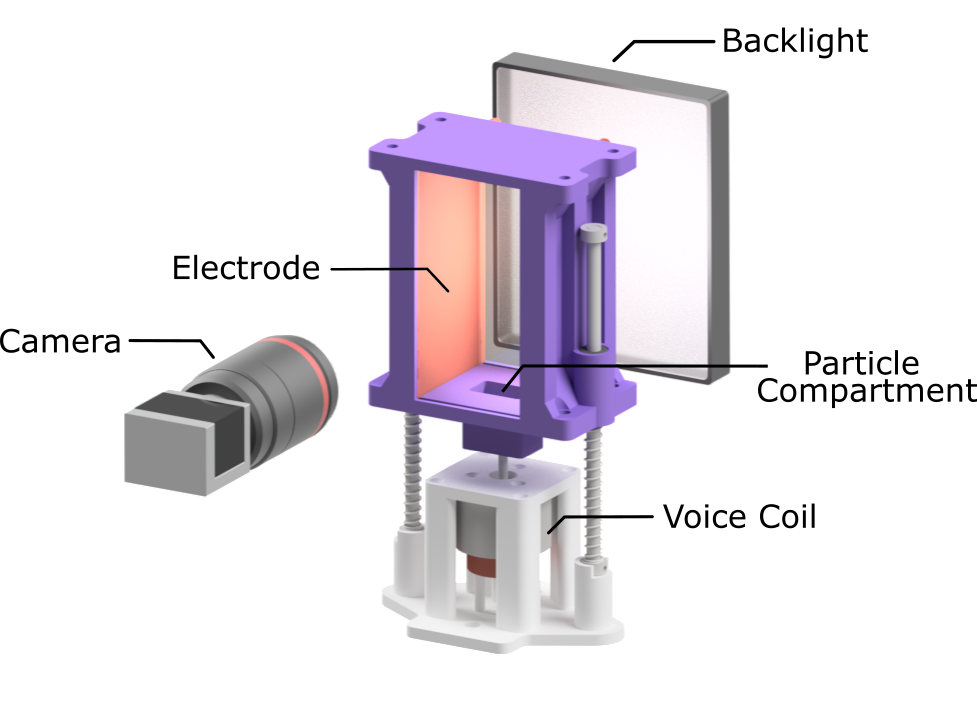}
      \caption{Sketch of the basic setup of the drop tower experiments. Dust aggregates within the particle compartment are charged by collisions. These are induced by vibrations as the voice coil is active. During the microgravity phase particles are released from the compartment. A high voltage can be applied to the electrodes of a plate capacitor (one electrode visible here). }
      \label{fig:aufbau}
   \end{figure}

The basic setup of the experiment is shown in fig. \ref{fig:aufbau}.
Two of the sides of the experiment chamber are glass to allow an observation of back-illuminated particles. The other two sides are copper electrodes which serve as a plate capacitor to provide an electric field.

The whole chamber including the aggregates are vibrated for 15 minutes before launch into microgravity. These vibrations are more gentle than the vibrations during aggregate production, but particles collide frequently here.
After the launch of the experiment in the drop tower, the particles move within the chamber in microgravity for about 9\,s. During this time, the chamber can be agitated further to induce particle motion for particles close to the wall and an electric field up to 160\,kV/m can be applied to the electrodes. We worked at ambient conditions, i.e. normal air at a pressure of about 1\,bar. Humidity was not controlled, neither were the particles pre-processed, e.g. heated to reduce their water content.

\subsection{Observations}

\begin{figure}
   \centering
      \includegraphics[width = \linewidth]{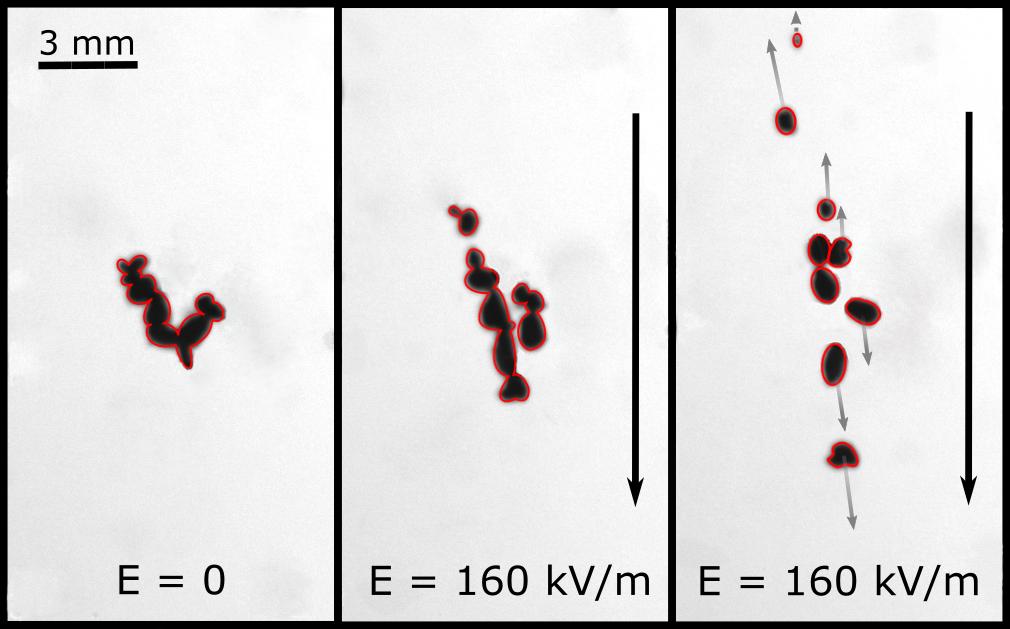}
      \caption{Chain of aggregates within the electric field of a plate capacitor. The outer aggregates charge and are subsequently removed, while the inner aggregates remain neutral. Arrows are introduced to visualize the direction of motion.}
      \label{fig:chain}
   \end{figure}

\begin{figure}
   \centering
      \includegraphics[width = \linewidth]{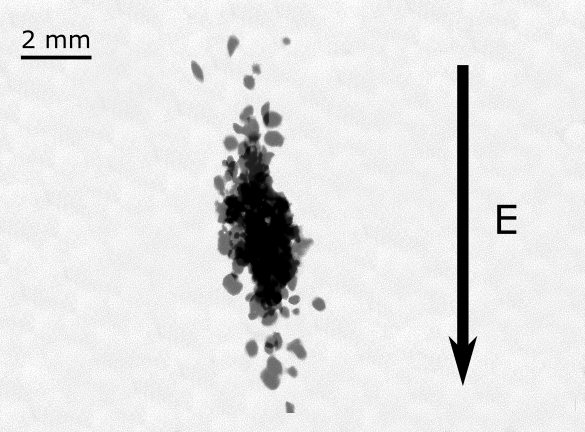}
      \caption{Erosion of a larger cluster of aggregates after the electric field was turned on; shown is a superposition of 4 images. The gray aggregates have been ripped off by the field and move upward and downward along the field lines. The black center part shows a temporarily stable core, but the cluster dissolves completely, eventually.}
      \label{fig:bigcluster}
   \end{figure}

During the microgravity phase we did observe the motion of aggregates and clusters of aggregates. These individual aggregates and clusters were parts of the original sample that broke into smaller units after launch. We did not observe a significant reformation of clusters from aggregates in collisions and do not consider collisions within the scope of this paper. However, the behavior of aggregates and clusters was quite systematic once the electric field was turned on. We only give two examples in fig. \ref{fig:chain} and \ref{fig:bigcluster} described below, but we note that the chamber was rather crowded with aggregates and all aggregates behaved like this. It can be summarized as follows:

\begin{itemize}
    \item Individual aggregates essentially move on straight trajectories, evidently carrying no significant amount of net charge.
    \item Clusters of aggregates disassamble by shedding aggregates from both ends with respect to the electric field direction. These aggregates are evidently charged and move further on to one of the electrodes. An example of a small chain dissolving is shown as an image sequence in fig. \ref{fig:chain}. Fig. \ref{fig:bigcluster} shows the erosion of a larger cluster. Here, the superposition of 4 images is shown.
    \item The center of mass of the remaining inner parts of an eroding cluster moves on a straight trajectory, i.e. it does not bear any net charge.
    \item By this procedure, aggregates are continuously detached until the whole cluster is gone or a small unit remains, which again has no net charge.
\end{itemize}

\section{Discussion}

The observed behavior is different from previous, similar experiments with monolithic, spherical sub-mm basalt and glass particles \citep{Steinpilz2020a, Steinpilz2020b, Jungmann2021c, Jungmann2022x}. In those earlier works, the samples consisted of insulating material, the vibrations would create charge patches across the particle surface that add up to a net charge and last for timescales longer than the duration of the experiment. After the initial vibrations, those particles formed clusters and kept grains that were net charged with both polarities. Applying an electric field did not disassemble these clusters. 

\begin{figure}
   \centering
      \includegraphics[width=\columnwidth]{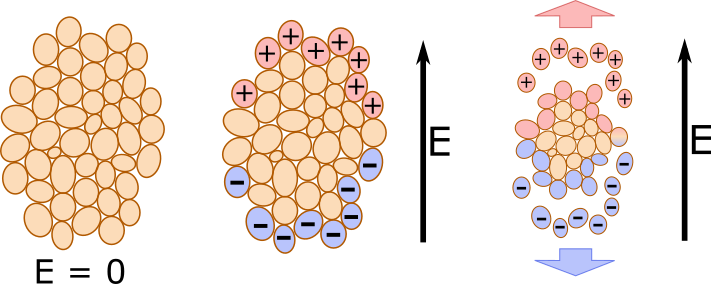}
      \caption{Pinciple idea of the electrostatic disassembly. left: stability without electric field; center: charge separation with external field; right: particle loss if forces are large enough. Once grains are fully detached, the field acting on the remaining aggregate is reestablished and erosion continues.}
      \label{fig:how_it_works}
   \end{figure}

Now, the sample is made up of µm-sized dust grains. The MGS aggregates contain mostly dielectric components and therefore isolators, but they are porous resulting in significant inner surface areas which invite surface water. With this in mind, all observations can be explained -- and only be explained satisfyingly so far -- if the aggregates are conductive. A sketch of a respective model is shown in fig. \ref{fig:how_it_works}.

If grains are conductive, they discharge during the launch phase of the experiment. Therefore, the center of mass of single aggregates and clusters does not react to an electric field and if particles are stable, they move on unbothered by the field. However, there is an internal charge separation. As conductors cannot have internal electric fields, positive surface charges face the negative electrode and vice versa as outlined in the center of fig. \ref{fig:how_it_works}. 

This exerts a tensile stress on clusters and aggregates. For a single aggregate, the tensile strength is strong enough to withstand this stress. However, the clusters consist of weakly bound aggregates and they break apart. As it is most likely that the outer aggregates have fewer bonds than the inner ones, they are bound most weakly and are released as soon as the separated charge is strong enough to provide the necessary force (fig. \ref{fig:how_it_works} right). The inner part that was field free is then again of net charge zero. The next layer of aggregates becomes charged as soon as the layer above has moved far enough so that the original external field can build up again, as indicated in fig. \ref{fig:how_it_works} (right). Then the next layer can be eroded and so on.

This disruption model is simple and very clear. The erosion of a layer of dust in external electric fields is actually not new and suggested / observed for dust and sand on a surface \citep{Onyeagusi2023a, Renno2008, Wang2016, Hirata2022, Kimura2022}. Here, electric fields for example within Earth's atmosphere, on asteroids or in experiments can lift grains from the surface. 

Our extension of this idea is the observation and proof that this concept works especially well for weakly bound clusters as they are supposed to form during planet formation, especially in the absence of gravity.

\subsection{Conductivity measurements}

\begin{figure}[H]
   \centering
      \includegraphics[width=\columnwidth]{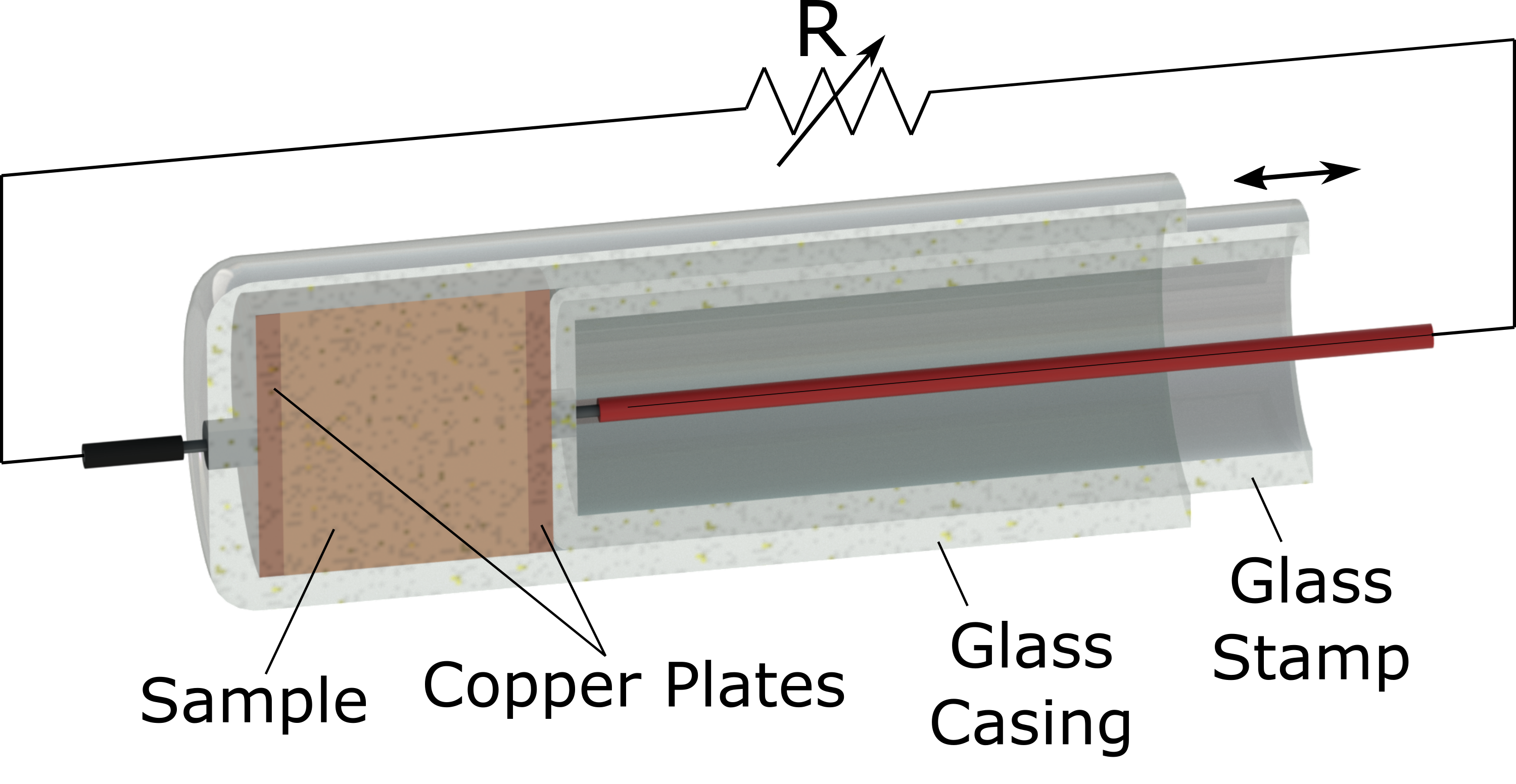}
      \caption{Setup for the measurement of the specific resistance of dust samples.}
      \label{fig:aufbau2}
   \end{figure}

The final proof for this model would be a measurement of the electric conductivity of the MGS sample. We therefore built a small setup to measure the conductivity of the sample in hindsight.
Fig. \ref{fig:aufbau2} shows this setup.
It measures the resistance in a simple cylindrical geometry in order to easily obtain a specific resistance. The dust sample is placed within a glass cylinder. The bottom of the cylinder is a copper electrode. A second inner glass cylinder with an electrode acts as a piston within the outer glass cylinder and compresses the sample. For the proof of concept of this paper, we do not quantify the dependence of the resistance on confining pressure, but the sample is manually compressed to have contacts along the whole top and bottom electrode. The length of the sample filling $l$ depends on the amount of material used. The cross section of the embedding cylinder is $A_c = 104 \mathrm{\,mm^2}$. The resistance $R$ throughout the sample is measured by an electrometer. In total, this gives a specific resistance $\rho$ defined as
\begin{equation}
    \rho = R \frac{A_c}{l}.
    \label{eq:resistance}
\end{equation}
For $l = 10 \mathrm{\,mm}$ we measure an average resistance of $R = 0.3 \mathrm{\,T\Omega}$ or $\rho = 3 \cdot 10^{15} \mathrm{\,\Omega \frac{mm^2}{m}}$.
The examined dust is actually not a very good conductor compared e.g. to a typical specific resistance of copper of $\rho_{Cu}=1.68\cdot 10^{-2} \mathrm{\,\Omega\frac{mm^2}{m}}$ \citep{Matula1979}.

\subsection{Charging time}

Despite its high resistivity, charging of the dust might still proceed on timescales short within the framework of the drop tower experiments. Dust aggregates would behave like a capacitor with resistance $R_a$ and capacitance $C_a$. The characteristic timescale would then be $\tau = R_aC_a$. To estimate the order of magnitude of this charging time, we approximate a dust aggregate as a cubic particle of size $d$. Then the resistance can be calculated using eq. \ref{eq:resistance} with $l=d$ and $A_c = A_a = d^2$. The capacitance can be estimated from a plate capacitor geometry or $C = \epsilon_0 \epsilon_r \frac{A_a}{d}$.
This gives a timescale of 
\begin{equation}
   \tau = R_aC_a = \rho \epsilon_0 \epsilon_r,
\end{equation}
which is actually independent of the particle size.

With the measured value for $\rho$, $\epsilon_0 = 8.8 \cdot 10^{-12} \mathrm{\frac{C}{V m}}$, and assuming $\epsilon_r = 2.5$ for silicate powder, this gives $\tau \sim 0.07 \mathrm{\,s}$.
This is \textbf{slightly faster than} the drop tower timescales. 
In fact, this timescale nicely fits to an earlier experiment by \citet{Onyeagusi2023a} where basalt was used in a similar setup. Here, it could be resolved on a sub-second timescale how aggregates recharged until being repelled from an electrode. 
We conclude that the aggregates charge on a timescale comparable to the camera resolution and can be considered to be conductive in the framework of our experiments, supporting the simple model outlined above.

\subsection{Threshold fields}

As outlined, the essentially complete destruction of all clusters implies that much smaller electric fields than the 160\,kV/m applied here are sufficient to disassemble the clusters of aggregates. The minimum fields needed for erosion to set in can be estimated from a force balance. In fact, the relevant balance was already used for basalt aggregates by \cite{Onyeagusi2023a} and referring to their work, we outline this here again.  

As the outer aggregates in a cluster are bound the least and are sequentially released, the force balance on individual aggregates is important. On one side, we do have the Coulomb force as the aggregates charge. On the other side, there is the adhesive force to the other aggregates. If both forces are the same in absolute terms, we are at the threshold of erosion. 

\cite{Onyeagusi2023a} showed that the contact forces acting on the constituent dust grains within the aggregates can be estimated based on the JKR adhesion theory. This theory describes the adhesive forces by taking into account the acting surface forces through the material specific surface energy. Instead of a smooth spherical surface, these aggregates are rough and grainy. Taking this into consideration, the adhesive force between single grains has to be multiplied by the number of contacts $N$ \citep{Johnson1971}, i.e.
\begin{equation}
F_{ad} = 3/2 \pi  \gamma r_d  N
\end{equation}
with dust grain radius $r_d$ and surface energy $\gamma$.
As particle size we take the typical $r_d = 0.5 \, \rm \mu m$ for protoplanetary dust.
The underlying assumption of our work is a weakly bound pebble cluster. Therefore, the number of contacts will not be large. Following \cite{Onyeagusi2023a} it is somewhat larger than 1. As this is only an order of magnitude estimate we somewhat arbitrarily assume $N=10$.
We leave $\gamma$ open for the moment.

Now, this adhesive force has to be overcome by the electrostatic pull
$F_{el} = q E$ with charge $q$ and electric field strength $E$.
In our model, the cluster and the aggregates within are (plate) capacitors.
Therefore, their charge density in an electric field in equilibrium after charging is
\begin{equation}
    \frac{q}{A_a} = \epsilon_0 E
\end{equation}
with aggregate surface $A_a$. For simplicity reasons, we again take $A=d^2$ as surface area, where $d=\rm 1 \, mm$ is the aggregate size assumed for particles at the bouncing barrier.
Now with this resulting maximum charge, the electrostatic force becomes
\begin{equation}
    F_{el} = A_a \epsilon_0 E^2
\end{equation}
and the threshold condition with $F_{el} = F_{ad}$ gives an electric field of
\begin{equation}
     E = \sqrt{ \frac{3 \pi  \gamma  r_d  N}{2 d^2 \epsilon_0}}.
\end{equation}
Introducing numeric values we get
$E \sim 10^{5} \, \rm V/m$ with a surface energy of $\gamma = \rm 0.01 \, J/m^2$. This is on the order of the field applied in the drop tower experiments. The laboratory setting is similar to the protoplanetary disk setting, but the required threshold field in the drop tower might be smaller than applied due to the different conditions. Normal pressure and humidity in the lab experiments can lower the surface energy by a factor of 10 compared to vacuum conditions in disks \citep{Pillich2021, Kimura2015, Steinpilz2019}. 
So while all quantities can be discussed and lowered or increased to some extent, calculations with $\gamma = \rm 0.01 \, J/m^2$ for disks and the experiments presented here are at least consistent.
In wind tunnel experiments, \citep{Schoenau2023} found that the erosion threshold for dust aggregates and solid spherical particles of similar size are comparable. We therefore do not expect the surface roughness to be of substantial influence in this setting.

\subsection{Electric fields in protoplanetary disks}

The importance of this new disassembly mechanism depends on the electric field, i.e. if protoplanetary disks harbor electric fields on the order of $\mathrm{10^5} \rm \,V/m$ or not.
Generally speaking, disks are very similar to atmospheres and most processes acting in planetary atmospheres can be expected to occur there as well. In planetary atmospheres, as for example on Earth, it is well known that particles like ice in a thunderstorm or dust in a volcanic eruption do charge \citep{Mason1988, Cimarelli2022}. The basic mechanism behind this is tribocharging \citep{Cimarelli2014, Shaw1926}. That means that whenever two particles collide, they exchange electric charge. This will not be different in protoplanetary disks. In fact, there have been quite a number of works in recent years that showed that tribocharging can fill a gap in the stages of planet formation, namely growing from the bouncing barrier at mm-size to larger clusters that are prone to concentration and gravitational instability \citep{Steinpilz2020a, Teiser2021, Jungmann2021c, Becker2022}.

In most collision experiments where small and large grains are involved, the small grains charge negatively while the large grains charge positively \citep{Waitukaitis2014, Lacks2011, Forward2009}. If grains are subsequently sorted by size, e.g. due to sedimentation, charge separation occurs on large scales and high electric fields can result in breakdown of the gas and, eventually, lightning \citep{Mason1988, Cimarelli2022}. 

For protoplanetary disks, lightning is mostly discussed in the context of chondrule formation, i.e. providing a heat source for flash heating \citep{Desch2000, Muranushi2010, Nuth2012, Kaneko2023}. 
Size sorting easily occurs in protoplanetary disks as well while grains sediment and drift with size dependent velocities \citep{Weidenschilling1977}. In fact, some size sorting is still conserved in meteorites \citep{Hughes1978, Kuebler1999, Metzler2019}. So one might expect large fields to be present in protoplanetary disks quite naturally.
There is an overwhelming number of works in support suggesting the generation of strong electric fields partially up to the point of lightning \citep{Balduin2023, Okuzumi2019, Inutsuka2005, Muranushi2012, Kaneko2023, Desch2000, Muranushi2010, Johansen2018}.

To put the necessary breakdown fields into context: In the well known case of Earth's atmosphere, 3\,MV/m are needed to generate lightning \citep{Melnik1998}. Even on Mars with a pressure close to the Paschen minimum, still about 25\,kV/m are needed for breakdown \citep{Kok2009}. 
Therefore, if conditions for lightning are met, the threshold for electrostatic erosion is also met and planetesimals or pebble piles should be prone to erosion in such regions.

\subsection{Conductivity of protoplanetary matter}

Electric conductivity might not be the first thing that comes to mind while thinking about planetesimals, but for the erosion process here, it is important.
We know from experiments with large monolithic (sub-mm) basalt and glass particles that they are bad conductors. Charge can be maintained on them for long timescales \citep{Onyeagusi2022, Becker2022}. This depends on the ambient conditions, e.g. long term experiments under vacuum show discharge timescales of months (Steinpilz, personal communication). Therefore, clusters of charged particles are supposed to form which in turn suggests that charge separation and lightning can occur.

In contrast, the MGS dust used here is somewhat conductive. So how does it fit together that dielectric, obviously bad conducting large grains tribocharge, form clusters or separate to produce lightning, while we discuss the disassembly of slightly conductive grains of dielectric composition?

We can identify at least three factors that are relevant in this context.
First, water on the surface plays a big role for tribocharging \citep{Jungmann2022x, Lee2018}. Dust aggregates have a large internal surface and therefore an increased influence of water which can raise the conductivity. This might be more of an issue for the experiments. In protoplanetary disks, at low ambient pressure, pebble pile planetesimals are not supposed to bear much water beyond a monolayer \citep{Pillich2021}. Particles still charge in collisions \citep{Becker2022}. This issue is unresolved.

Second, the composition of dust in protoplanetary disks is not just silicates. At high temperatures, for example, metallic iron is formed \citep{Bogdan2023}. The immediate witnesses of iron in protoplanetary disks are iron meteorites. But metallic iron also finds its way into chondrites \citep{Rindlisbacher2021}.
Surface water might be present or not depending on the formation region of the body and its further history.

Third, the timescales play an important role. If, for example, the lower water content made clusters slightly conductive, but charge separation timescales were long with respect to contact times or particle separation to produce large scale fields, then tribocharging might occur while a planetesimal drifts into this region and is slowly charged and destroyed.
Unfortunately, none of this can currently be constrained any better.

\section{Conclusions}

Pebble pile planetesimals have to be real survival artists with their fragile nature. What we show here is that the dawn of an upcoming thunderstorm in protoplanetary disks might seal the fate of a planetesimal. Less visually speaking, strong electric fields can disassemble pebble piles consisting of only somewhat electrically conductive matter easily.
This erosion process provides pebbles, which might be observable and hint to electric activity in protoplanetary disks.

Constraints from this study would be that planetesimals are safe, if they are not pebble piles, they are safe if they are not conductive and they are safe if the electric fields are too low. However, in models where planetesimals are pebble piles with large fractions of metallic iron or conductive surface layers passing through a region with stronger electric fields, they might dissolve.

\section*{Acknowledgements}

This project is supported by DLR Space Administration with funds provided by the Federal Ministry for Economic Affairs and Climate Action (BMWK) under grant number DLR 50 WM 2142. 
T.B. is funded by the European
Union’s Horizon 2020 research and innovation
program under grant agreement No 101004052. We acknowledge support by the Open Access Publication Fund of the University of Duisburg-Essen. We thank G.Völke for running the conductivity measurements. We also thank Mihály Horányi for the helpful comments as a reviewer.\\

This is a pre-copyedited, author-produced PDF of an article accepted for publication in \textit{Monthly Notices of the Royal Astronomical Society} following peer review. The version of record is available online at: \url{https://doi.org/10.1093/mnras/stae599}.

\section*{Data Availability}

The data that support the findings of this study are available from the corresponding author, F.C.O., upon reasonable request.
 



\bibliographystyle{mnras}
\bibliography{bib} 








\bsp	
\label{lastpage}
\end{document}